\documentclass[twocolumn]{article}
\usepackage[utf8]{inputenc}
\usepackage[english]{babel}
\usepackage{natbib}
\usepackage{graphicx, caption}
\usepackage{authblk}
\usepackage{textcomp}

\usepackage{multicol}
\usepackage[margin=0.6in]{geometry}

\immediate\write18{texcount -char -tex -sum ./main.tex > ./wordcount.tex}

\title{Periodicity in fields of elongating dunes}
\author[1]{C. Gadal}
\author[1]{C. Narteau}
\author[3]{S. Courrech du Pont}
\author[1]{O. Rozier}
\author[2]{P. Claudin}
\affil[1]{Institut de Physique du Globe de Paris, UMR 7154 CNRS, Université de Paris, 75005 Paris, France}
\affil[2]{Laboratoire Matière et Systèmes Complexes, UMR 7057 CNRS, Université de Paris, 75205 Paris, France}
\affil[3]{Physique et Mécanique des Milieux Hétérogènes, UMR 7636 CNRS, ESPCI Paris and PSL Research University–Sorbonne
Université–Université de Paris, 75005 Paris, France}

\date{}

\begin{document}

\twocolumn[
\maketitle

\begin{abstract}
Dune fields are commonly associated with periodic patterns that are among the most recognizable landscapes on Earth and other planetary bodies. However, in zones of limited
sediment supply, where periodic dunes elongate and align in the direction of the resultant sand flux, there has been no attempt to explain the emergence of such a regular pattern. Here, we show, by means of numerical simulations, that the elongation growth mechanism does not produce a pattern with a specific wavelength. Periodic elongating dunes appear to be a juxtaposition of individual structures, the arrangement of which is due to regular landforms at the border of the field acting as boundary conditions. This includes, among others, dune patterns resulting from bed instability, or the crestline reorganization induced by dune migration. The wavelength selection in fields of elongating dunes therefore reflects the interdependence of dune patterns over the course of their evolution.
\end{abstract}
\vskip 0.7cm
]

\begin{figure*}[h]
\centering
	\includegraphics{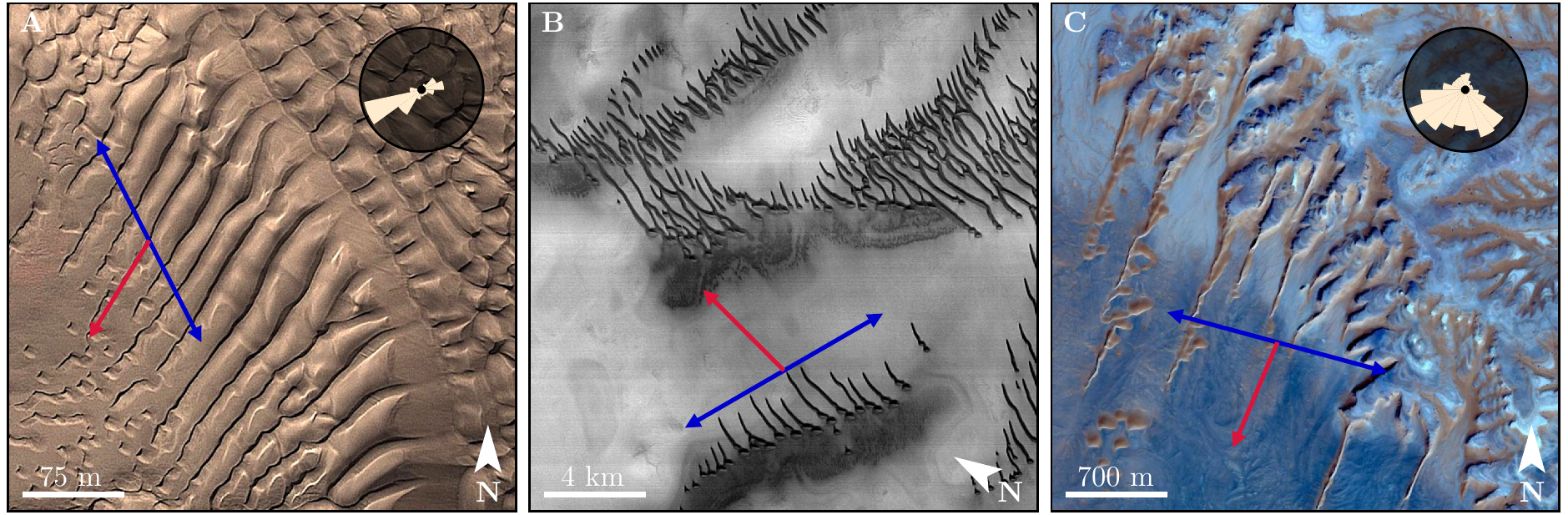}
	\caption{\textbf{Periodic dune patterns.} \textbf{A}:~Elongating dunes at bottom of an avalanche slope, Taklamakan desert, China ($37^\circ 45'$~N, $82^\circ 36'$~E). \textbf{B}:~Elongating dunes on Mars, Scandia Cavi ($78^\circ 04'$~N, $151^\circ 10'$~W). \textbf{C}:~Elongating dunes at bottom of an erosion pattern, Rub' Al Khali Desert, Saudi Arabia, ($10^\circ 00'$~N, $45^\circ 40'$~E). Arrows show the dune orientations predicted from wind data on nonerodible bed (red) and sand layer (blue), following \protect \citet{Cour14}, except for Mars (\textbf{B}), where dune orientations were inferred from dune crests on satellite image. Insets show sand flux rose diagrams. Satellite images: Google\texttrademark, Maxar Technologies (Colorado, USA), NASA/JPL/University of Arizona.}
	\label{Figure_terrain}
\end{figure*}

\section*{Introduction}
Systematically highlighted by aerial photographs and satellite images, periodic geomorphological features have revealed the presence of dunes and atmospheric flows on planetary bodies including Earth, Mars, Titan, Pluto, and comets \citep{Mcke79, Cutt73, Ward85, Lore06, Bour10, bLore14, Luca14, Jia17, Dini17, Telf18}. The origin of these periodic patterns is a central issue in dune physics and planetary sciences, along with the diversity of dune shapes, sizes, and orientations \citep{Wass83, Rubi87, Andr09}.

Under multidirectional flows, many dune fields with low sand availability exhibit linear dunes that extend for kilometers on nonerodible beds. These dunes result from an elongation growth mechanism leading to deposition at the dune tip of sediment transported along the crest under the action of reversing winds \citep{Cour14,Gao15}. While these elongating dunes can exist as isolated objects \citep{Luca15}, most fields of elongating dunes display a periodicity (Fig.~\ref{Figure_terrain}). The origin of this periodicity has not yet been investigated or compared to pattern formation in zones where dunes can grow in height with full sediment availability. A related point is thus how populations of dunes in transport- and sediment-limited conditions are independent based on current understanding of the emergence of dune wavelengths.

Periodic dune patterns are typically linked to the development of incipient bed forms due to an instability mechanism, which selects a characteristic length scale from the competition between destabilizing and stabilizing processes. The flat bed instability occurs in transport-limited conditions, where dunes arise from the interactions among turbulent flow, sediment transport, and topography at a wavelength $\lambda_{\rm max}$, for which the growth rate is maximal \citep{Char13}. AAnother instability occurs on transverse dunes migrating on a nonerodible bed. They break into a set of periodic barchan dunes with a size proportional to the height of the initial transverse dune \citep{Reff10, Niiy10, Part11, Guig13}. The collective dynamics of individual bed forms can also govern the development of patterns. In sediment-limited environments, these individual structures take the shape of barchans, domes, or star dunes according to the wind regime \citep{Hers04,Zhan12,Badd18,Gao18}. Whatever their shapes, as soon as they form populations, dunes interact by mass exchange induced either by collisions or sand flux. These interactions generate a higher level of organization, with a characteristic size, frequently observed along the sand flow paths in dune corridors, chains, and clusters \citep{Elbe08, Worm13, Geno13, Geno13b}. In all dune fields, independent of the origin of the pattern, nonlinear interactions also lead to an increase in dune amplitude and wavelength in space and time  \citep{Ewin10,Vala11,Gao15a}. While the smallest wavelengths are intrinsically associated with incipient bed forms, the coarsening process is restricted by boundary conditions, which can ultimately set the maximum length scales of the dune pattern.

By means of numerical simulations, we studied two independent configurations that produce periodic elongating dunes. We show that the spatial organization of these dunes is controlled by the geomorphological patterns from which they develop. The elongation growth mechanism does not itself select a characteristic wavelength but can be expressed from preexisting structures while retaining some of their properties. The periodic pattern is therefore inherited from past environments or specific boundary conditions upstream of the sand flow paths.


\section*{Methods}

\begin{figure}
	\centering
	\includegraphics{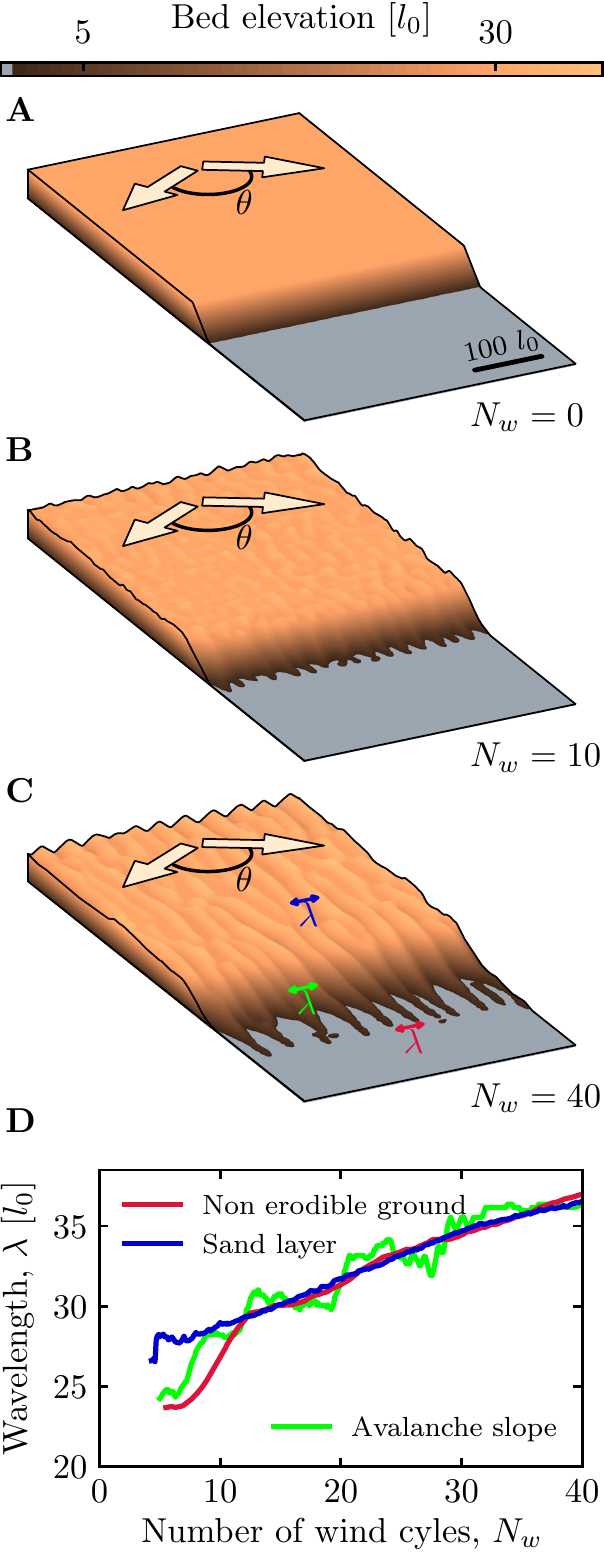}
	\caption{\textbf{Dunes elongating at border of a sediment layer}. \textbf{A-C}: Evolution of sediment bed in model. Nonerodible bed is shown in gray. \textbf{D}: Wavelengths of pattern on sediment layer (blue), on avalanche slope (green) and on nonerodible (red) bed as function of time.}
	\label{Figure_lit_plat}
\end{figure}

We used a cellular automaton dune model that accounts for feedback mechanisms between the flow and the evolving bed topography \citep{Rozi14}. This approach has been shown to be efficient in modeling dunes in multidirectional wind regimes and is able to reproduce and quantify superimposed bed forms in dune fields \citep{Zhan12, Lu17}. The model parameters were the same as in \citet{Nart09},  so that the model length and time units $l_{0}$ and $t_{0}$ were $\simeq 0.5$~m and $\simeq 10^{-3}$~yr, respectively. Based on field observations, we implemented two specific configurations to generate periodic elongating dunes. In both cases, we considered a bidirectional wind regime with a divergence angle $\theta=120^\circ$ and a constant wind strength (saturated flux over a flat sand bed $\simeq 0.23~l_{0}^{2}~t_{0}^{-1}$). However, the duration of the winds was allowed to differ, and the transport ratio $N$ was defined as the ratio between the time spent in the primary and secondary winds over a wind cycle $T_{\rm w}$ \citep{Rubi87}. Autocorrelation of the bed elevation was used to measure the elongation rate and the wavelength of the dune patterns.

In the first configuration, we focused on the development of dune patterns at the interface between a sediment layer of height 30l0 and a nonerodible bed. An avalanche slope made the transition between the two (inset of Fig.~\ref{Figure_lit_plat}). We used the simplest case of winds of equal duration ($N=1$, $T_{\rm w} = 100~t_{0}$) with a resultant sand transport perpendicular to the avalanche slope. All dune orientations were aligned with the transport direction, and there was no lateral dune migration.

Along this direction, the output sand flux was lost (i.e., open conditions). Perpendicularly, the cellular space was made periodic to mimic an infinitely wide system.

In the second configuration, a sand bar of height $H$ was placed over a nonerodible bed. We studied its migration when subjected to an asymmetric regime ($N = 3$, $T_{\rm w} = 40~t_{0}$) for which the dominant wind was perpendicular to the bar orientation (Fig.~\ref{Figure_barre}A). Along this direction, we kept open the output sand flux condition, preserving the periodicity in the perpendicular direction. As in a unidirectional wind regime, the bar migrates and can destabilize due to the lateral mass redistribution. Furthermore, in the presence of a secondary wind, elongating dunes can develop.

Finally, we interpret field examples using the method described in \citet{Cour14}. The predicted dune orientations were calculated from wind data provided by the
ERA-Interim project or by local wind towers (Fig.~\ref{Figure_terrain}B). For the Martian example, they were measured from the dune crests on the satellite images (Fig.~\ref{Figure_terrain}C).

\begin{figure*}
	\centering
	\includegraphics{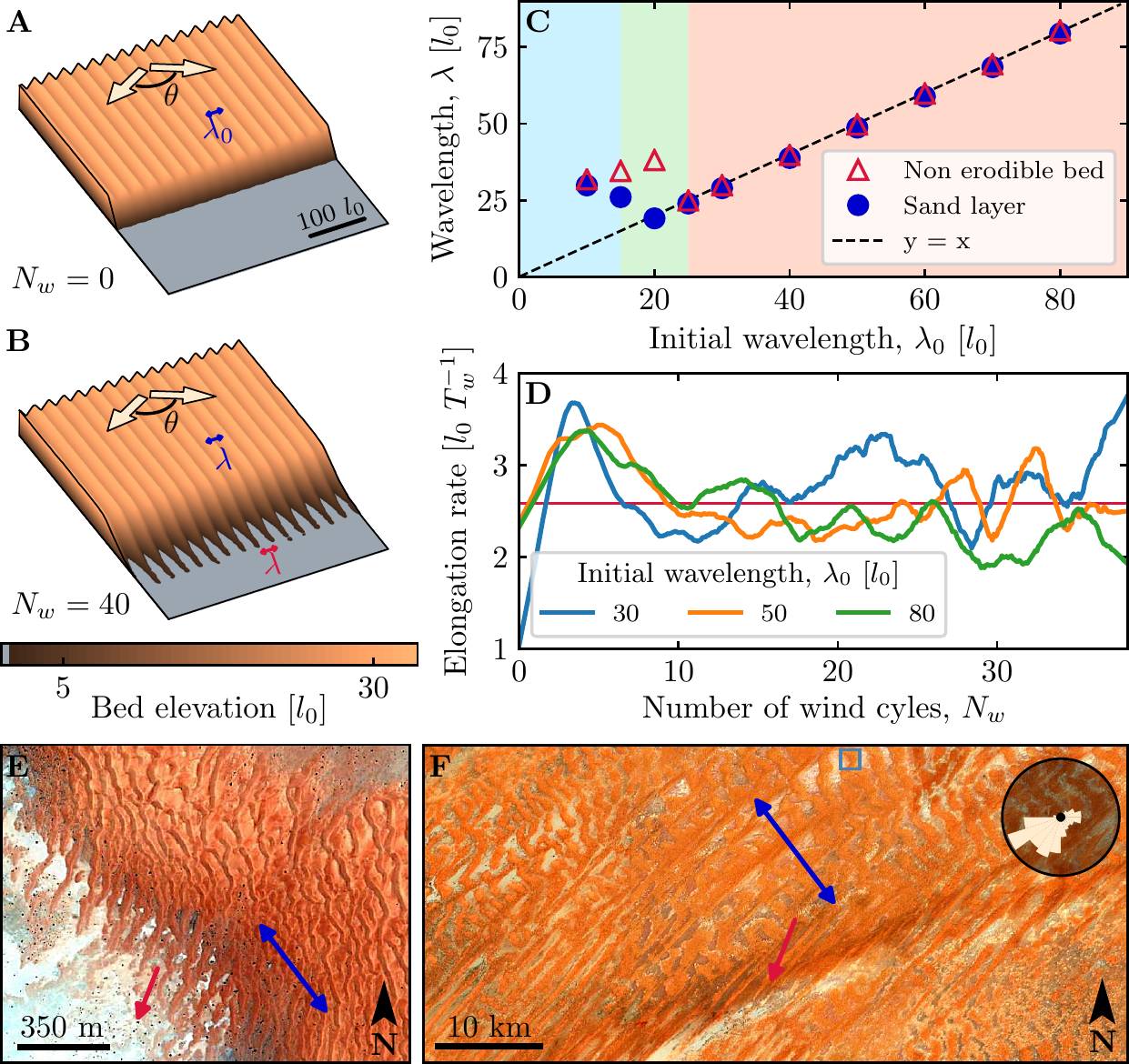}
	\caption{\textbf{Dunes elongating at border of sinusoidal sediment layer}. \textbf{A-B}: Evolution of sediment bed in model for initial wavelength $\lambda_0=30\;l_{0}$ ($l_{0}-$ model length unit). \textbf{C}: Wavelengths of pattern on sediment bed (blue) and on nonerodible bed (red) after $40$ wind cycles as function of imposed wavelength $\lambda_0$. \textbf{D}: Elongation rate of dunes as function of time for different imposed wavelengths. Red line shows average value. \textbf{E/F}: Field examples at small (\textbf{E}) and large (\textbf{F}) scales in western Sahara, Mauritania ($18^\circ 11'$~N, $14^\circ 30'$~W). Blue square is location of \textbf{E} on \textbf{F}. Arrows show dune orientations predicted from wind data on nonerodible bed (red) and sand layer (blue) following \protect \citet{Cour14}. Insets show sand flux rose diagrams. Satellite images: Google\texttrademark, Maxar Technologies (Colorado, USA), Landsat/Copernicus}
	\label{Figure_lit_pert}
\end{figure*}

\section*{Results}

Figures~\ref{Figure_lit_plat}A-C show the evolution of the bed forms at the border of the sediment layer. Incipient dunes develop from the flat bed instability on the sediment layer and the avalanche slope, and also elongate at the bottom of this lee face. Both dune types exhibit a regular pattern with well-defined wavelengths that connect with each other, as well as the same orientation due to the specific wind regime $N = 1$ (Fig.~\ref{Figure_lit_plat}C). On the sediment layer, the dune wavelength starts from that of the most unstable mode associated with the bidirectional wind regime \citep{Gadal19}.  On the avalanche slope, the pattern wavelength starts at a slightly lower value, as does the one on the nonerodible ground. After a few wind cycles, both patterns connect through the avalanche slope, as the wavelengths become identical and then increase due to coarsening (Fig.~\ref{Figure_lit_plat}D). The periodicity of the elongating dunes then appears to be controlled by that of the dunes on the sand layer. In addition, we observe that the pattern of the elongating dunes is directly impacted by the defects of the sediment layer pattern when the dunes reach the avalanche slope.

To test this control mechanism, we started simulations from a sediment layer with a sinusoidal elevation profile perpendicular to the resultant sand flux (Fig.~\ref{Figure_lit_pert}A-B). Figure~\ref{Figure_lit_pert}C shows the wavelength of the dune patterns on both sides of the avalanche slope after $40$ wind cycles with respect to the initial wavelength~($\lambda_{\rm 0}$). For $\lambda_{\rm 0}$ larger than $\lambda_{\rm max}$, there is perfect agreement between all these wavelengths (red part of Fig.~\ref{Figure_lit_pert}C). As these large wavelengths are unstable with respect to the flat bed instability, the initial pattern persists throughout the simulation, enforcing its periodicity onto the elongating dunes. For $\lambda_{\rm 0}$ smaller than $\lambda_{\rm max}$, however, the initial bed undulations are quickly replaced by those of the most unstable mode of the flat bed instability (blue part of Fig.~\ref{Figure_lit_pert}C). As coarsening occurs along the resultant flux direction, the wavelength of the elongating dunes is a bit larger than the one measured on the sediment layer.

Figure~\ref{Figure_lit_pert}D  shows the elongation rate of the dunes developing at the bottom of the avalanche slope as a function of time. These measurements were done for three large wavelengths (red part in Fig.~\ref{Figure_lit_pert}C). After a short transient time period of 5-10 wind cycles, all three elongation rates fluctuate around the same constant value. This duration corresponds to the time needed for the dunes at the base of the avalanche slope to reach the minimal size integrating the two winds. In the next phase, during which they elongate at a constant average rate, the dune tip periodically breaks, migrates, and disperses, explaining the observed variability. As no wavelength emerges faster than the others, the elongation mechanism does not select any length scale.

\begin{figure*}
	\centering
	\includegraphics{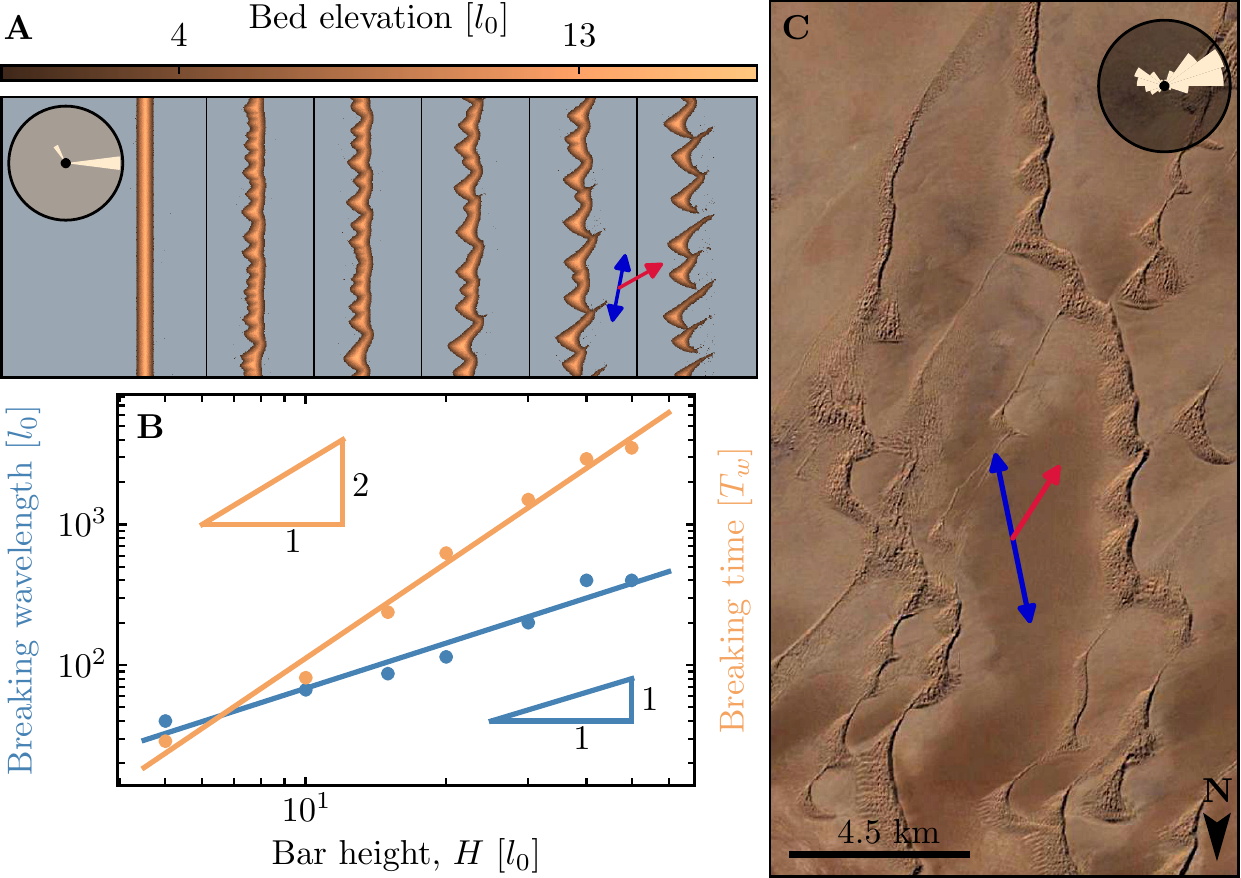}
	\caption{\textbf{Breakup of a sandbar in bidirectional wind regime}. \textbf{A}: Evolution of a sandbar during its migration (corresponding number of wind cyles: $N_{w} = [0,~150,~225,~335,~400,~580]$). \textbf{B}: Wavelength of dune pattern and migration time as function of the initial height of bar, measured at breaking. Points are numerical results and lines are best linear fits. \textbf{C}: Field example in the Kavir desert, Iran ($34^\circ 06'$~N, $53^\circ 40'$~E). Arrows show dune orientations predicted from wind data on nonerodible bed (red) and sand layer (blue) following \protect \citet{Cour14}. Insets show sand flux rose diagrams. Satellite images: Google\texttrademark, Maxar Technologies (Colorado, USA)}
	\label{Figure_barre}
\end{figure*}

Using the second configuration, Figure~\ref{Figure_barre}A shows a migrating sandbar that becomes sinuous and ultimately breaks up into a periodic set of
dunes. As in unidirectional wind regimes, the resulting wavelength of these dunes is determined by the height of the sandbar (Fig.~\ref{Figure_barre}). In bidirectional wind regimes, the crestline reorganization is not only related to the lateral redistribution of mass associated with the primary wind, but also to the superimposed bed forms that develop and migrate in an oblique direction with respect to the orientation of the sandbar \citep{Cour15, Lu17}. Ultimately, the resulting periodic dunes also elongate under the action of the reversing winds \citet{Cour14} (Fig.~\ref{Figure_barre}C), as in the experiments of \citet{Cour14}. In multidirectional flows, the breakup of preexisting dunes is therefore another source of periodicity in fields of elongating dunes (Fig.~\ref{Figure_terrain}B).

\section*{Discussion}

We identified two relevant length scales for dune growth in bidirectional wind regimes that are potential candidates for wavelength control in fields of elongating dunes. The first is the saturation length, associated with the spatial relaxation of transport in response to a bed perturbation \citep{Andr10}. The second is proportional to $\sqrt{QT_{w}}$, where $Q$ is the characteristic sand flux and $T_{w}$ is the duration of the wind cycle. As shown by \citet{Rozi19}, these two length scales control the minimal size and the morphology of isolated elongating dunes. Nevertheless, our numerical simulations suggest that the periodicity in fields of elongating dunes comes from boundary and initial conditions rather than from length scales inherent to the elongation mechanism. This is supported by the fact that the elongation rate is independent of the initial wavelength (Fig.~\ref{Figure_lit_pert}D).

We then recognized in this process the importance of the migration of dunes from areas of high to low sediment availability. When these dunes eventually fall into an avalanche slope in the transition zone, they impose a periodic modulation at its base, from which elongating dunes can develop (Fig.~\ref{Figure_terrain}A). Interestingly, avalanche processes, and especially sand spreading during granular flow, can also add a level of complexity. In our numerical simulations, high slip faces can act as a filter and prevent transmission of upstream dune patterns that have too short an amplitude or wavelength down to the base of the avalanche slope. As a result, there is a range of wavelengths for which information is lost in the avalanche slope (green part of Fig.~\ref{Figure_lit_pert}C).

In addition, superimposed bed forms also appear and develop directly on avalanche slopes, which are areas of loose sand similar to a flat bed for oblique winds. In doing so, they also allow the elongation of periodic dunes at the base. Finally, multidirectional wind regimes more complex than the simplest symmetric bidirectional case studied in Figures~\ref{Figure_lit_plat} and \ref{Figure_lit_pert} would induce an upstream dune orientation oblique to the avalanche slope. As a consequence, under natural conditions, we would expect additional geometric factors to be involved in the relation between the wavelength of the upstream bed forms and that of the elongating dunes.

Although the periodicity of many fields of elongating dunes appears to be governed by upstream dune patterns along the sand flow path, it can also be attributed to other spatially periodic landscapes. For example, channel formation and river erosion can lead to evenly spaced ridges and valleys (Fig.~\ref{Figure_terrain}C). This could naturally give rise to a new type of interaction between eolian and fluvial system in dryland environments \citep{Bull02}.

In modern sand seas, large numbers of extremely regular fields of elongating dunes have giant sizes, i.e., with a kilometer-scale spacing \citep{Andr09}. Although we mainly addressed small- and medium-sized dunes in our numerical simulations, our results suggest that boundary conditions determine the periodicity of elongating dunes at any scale. When the dunes at the border have reached their giant size, the wavelength of the elongating dunes must adapt accordingly, preserving the periodicity along the sand flow paths (see Figure~\ref{Figure_lit_pert}E-F). However, long-term interactions between elongating linear dunes and the resulting auto-organization require more investigation. In particular, dunes exchange sand by means of lateral fluxes, which make them expand longer \citep{Rozi19}. Whether this process acts as a stabilizing or a destabilizing mechanism for the pattern is an open question.

More generally, our results illustrate how the transmission of the pattern at the border of dune fields exerts a spatio-temporal control on the geomorphology of the landscape along the sand flow path. Since any change at the border could be traced farther downwind over time, the observed dune morphodynamics could provide information about upstream dune patterns or past environments. They introduce a long-term memory into the migrating dune system. The propagative aspect of the dynamics of the pattern is also apparent in the presence of an obstacle or in the elimination of defects. Because the density of defects is lower in fields of elongating dunes \citep{Day18}, periodic patterns emerge more easily, even if they only develop under specific boundary conditions. On Earth, but also on other planetary bodies, these dynamic aspects can now be treated more carefully at the scale of major sand seas to learn more about the long-term evolution of arid landscapes and the variety of dune field patterns.

\section*{Acknowledgment}
We acknowledge financial support from the UnivEarthS LabEx program of Sorbonne Paris Cité (grants ANR-10-LABX- 0023 and ANR-11-IDEX-0005–02) and the French National Research Agency (grant ANR-17-CE01–0014/SONO). Clément Narteau acknowledges support from the National Science Center of Poland (grant 2016/23/B/ST10/01700).

\bibliographystyle{apalike}
\bibliography{Biblio}

\begin{thebibliography}{}

\bibitem[Andreotti et~al., 2010]{Andr10}
Andreotti, B., Claudin, P., and Pouliquen, O. (2010).
\newblock Measurements of the aeolian sand transport saturation length.
\newblock {\em Geomorphology}, 123(3-4):343--348.

\bibitem[Andreotti et~al., 2009]{Andr09}
Andreotti, B., Fourriere, A., Ould-Kaddour, F., Murray, B., and Claudin, P.
  (2009).
\newblock Giant aeolian dune size determined by the average depth of the
  atmospheric boundary layer.
\newblock {\em Nature}, 457:1120.

\bibitem[Baddock et~al., 2018]{Badd18}
Baddock, M., Nield, J., and Wiggs, G. (2018).
\newblock Early-stage aeolian protodunes: Bedform development and sand
  transport dynamics.
\newblock {\em Earth Surface Processes and Landforms}, 43:339--346.

\bibitem[Bourke et~al., 2010]{Bour10}
Bourke, M.~C., Lancaster, N., Fenton, L.~K., Parteli, E., Zimbelman, J., and
  Radebaugh, J. (2010).
\newblock Extraterrestrial dunes: An introduction to the special issue on
  planetary dune systems.
\newblock {\em Geomorphology}, 121:1--14.

\bibitem[Bullard and Livingstone, 2002]{Bull02}
Bullard, J. and Livingstone, I. (2002).
\newblock Interactions between aeolian and fluvial systems in dryland
  environments.
\newblock {\em Area}, 34(1):8--16.

\bibitem[Charru et~al., 2013]{Char13}
Charru, F., Andreotti, B., and Claudin, P. (2013).
\newblock Sand ripples and dunes.
\newblock {\em Annual Review of Fluid Mechanics}, 45:469--493.

\bibitem[{Courrech du Pont}, 2015]{Cour15}
{Courrech du Pont}, S. (2015).
\newblock Dune morphodynamics.
\newblock {\em Comptes Rendus Physique}, 16:118--138.

\bibitem[{Courrech du Pont} et~al., 2014]{Cour14}
{Courrech du Pont}, S., Narteau, C., and Gao, X. (2014).
\newblock Two modes for dune orientation.
\newblock {\em Geology}, 42:743--746.

\bibitem[Cutts and Smith, 1973]{Cutt73}
Cutts, J. and Smith, R. (1973).
\newblock Eolian deposits and dunes on mars.
\newblock {\em Journal of Geophysical Research}, 78:4139--4154.

\bibitem[Day and Kocurek, 2018]{Day18}
Day, M. and Kocurek, G. (2018).
\newblock Pattern similarity across planetary dune fields.
\newblock {\em Geology}, 46:999--1002.

\bibitem[Diniega et~al., 2017]{Dini17}
Diniega, S., Kreslavsky, M., Radebaugh, J., Silvestro, S., Telfer, M., and
  Tirsch, D. (2017).
\newblock Our evolving understanding of aeolian bedforms, based on observation
  of dunes on different worlds.
\newblock {\em Aeolian research}, 26:5--27.

\bibitem[Elbelrhiti et~al., 2008]{Elbe08}
Elbelrhiti, H., Andreotti, B., and Claudin, P. (2008).
\newblock Barchan dune corridors: field characterization and investigation of
  control parameters.
\newblock {\em Journal of Geophysical Research}, 113(F02S15).

\bibitem[Ewing and Kocurek, 2010]{Ewin10}
Ewing, R. and Kocurek, G. (2010).
\newblock Aeolian dune-field pattern boundary conditions.
\newblock {\em Geomorphology}, 114:175--187.

\bibitem[Gadal et~al., 2019]{Gadal19}
Gadal, C., Narteau, C., {Courrech du Pont}, S., Rozier, O., and Claudin, P.
  (2019).
\newblock Incipient bedforms in a bidirectional wind regime.
\newblock {\em Journal of Fluid Mechanics}, 862:490--516.

\bibitem[Gao et~al., 2018]{Gao18}
Gao, X., Gadal, C., Rozier, O., and Narteau, C. (2018).
\newblock Morphodynamics of barchan and dome dunes under variable wind regimes.
\newblock {\em Geology}, 46:743--746.

\bibitem[Gao et~al., 2015a]{Gao15a}
Gao, X., Narteau, C., and Rozier, O. (2015a).
\newblock Development and steady states of transverse dunes: A numerical
  analysis of dune pattern coarsening and giant dunes.
\newblock {\em Journal of Geophysical Research: Earth Surface}, 120:2200--2219.

\bibitem[Gao et~al., 2015b]{Gao15}
Gao, X., Narteau, C., Rozier, O., and Courrech Du~Pont, S. (2015b).
\newblock Phase diagrams of dune shape and orientation depending on sand
  availability.
\newblock {\em Scientific reports}, 5:14677.

\bibitem[G{\'e}nois et~al., 2013a]{Geno13}
G{\'e}nois, M., Courrech Du~Pont, S., Hersen, P., and Gr{\'e}goire, G. (2013a).
\newblock An agent-based model of dune interactions produces the emergence of
  patterns in deserts.
\newblock {\em Geophysical Research Letters}, 40:3909--3914.

\bibitem[G{\'e}nois et~al., 2013b]{Geno13b}
G{\'e}nois, M., Courrech Du~Pont, S., Hersen, P., and Gr{\'e}goire, G. (2013b).
\newblock Spatial structuring and size selection as collective behaviours in an
  agent-based model for barchan fields.
\newblock {\em The European Physical Journal B}, 86(11):447.

\bibitem[Guignier et~al., 2013]{Guig13}
Guignier, L., Niiya, H., Nishimori, H., Lague, D., and Valance, A. (2013).
\newblock Sand dunes as migrating strings.
\newblock {\em Physical Review E}, 87:052206.

\bibitem[Hersen, 2004]{Hers04}
Hersen, P. (2004).
\newblock On the crescentic shape of barchan dunes.
\newblock {\em The European Physical Journal B-Condensed Matter and Complex
  Systems}, 37:507--514.

\bibitem[Jia et~al., 2017]{Jia17}
Jia, P., Andreotti, B., and Claudin, P. (2017).
\newblock Giant ripples on comet 67p/churyumov--gerasimenko sculpted by sunset
  thermal wind.
\newblock {\em Proceedings of the National Academy of Sciences},
  114:2509--2514.

\bibitem[Lorenz et~al., 2006]{Lore06}
Lorenz, R., Wall, S., Radebaugh, J., Boubin, G., Reffet, E., Janssen, M.,
  Stofan, E., Lopes, R., Kirk, R., Elachi, C., et~al. (2006).
\newblock The sand seas of titan: Cassini radar observations of longitudinal
  dunes.
\newblock {\em Science}, 312:724--727.

\bibitem[Lorenz and Zimbelman, 2014]{bLore14}
Lorenz, R.~D. and Zimbelman, J.~R. (2014).
\newblock {\em Dune worlds: How windblown sand shapes planetary landscapes}.
\newblock Springer Science \& Business Media.

\bibitem[L{\"u} et~al., 2017]{Lu17}
L{\"u}, P., Narteau, C., Dong, Z., Rozier, O., and {Courrech du Pont}, S.
  (2017).
\newblock Unravelling raked linear dunes to explain the coexistence of bedforms
  in complex dunefields.
\newblock {\em Nature Communications}, 8:14239.

\bibitem[Lucas et~al., 2015]{Luca15}
Lucas, A., Narteau, C., Rodriguez, S., Rozier, O., Callot, Y., Garcia, A., and
  Courrech~du Pont, S. (2015).
\newblock Sediment flux from the morphodynamics of elongating linear dunes.
\newblock {\em Geology}, 43:1027--1030.

\bibitem[Lucas et~al., 2014]{Luca14}
Lucas, A., Rodriguez, S., Narteau, C., Charnay, B., {Courrech du Pont}, S.,
  Tokano, T., Garcia, A., Thiriet, M., Hayes, A., Lorenz, R., et~al. (2014).
\newblock Growth mechanisms and dune orientation on titan.
\newblock {\em Geophysical Research Letter}, 41(17):6093--6100.

\bibitem[McKee, 1979]{Mcke79}
McKee, E. (1979).
\newblock Introduction to a study of global sand seas.
\newblock In {\em A study of global sand seas}, volume 1052, pages 1--19.
  Professional Paper.

\bibitem[Narteau et~al., 2009]{Nart09}
Narteau, C., Zhang, D., Rozier, O., and Claudin, P. (2009).
\newblock Setting the length and time scales of a cellular automaton dune model
  from the analysis of superimposed bed forms.
\newblock {\em Journal of Geophysical Research: Earth Surface}, 114:F03006.

\bibitem[Niiya and Nishimori, 2010]{Niiy10}
Niiya, H.and~Awazu, A. and Nishimori, H. (2010).
\newblock Three-dimensional dune skeleton model as a coupled dynamical system
  of two-dimensional cross sections.
\newblock {\em Journal of the Physical Society of Japan}, 79:063002.

\bibitem[Parteli et~al., 2011]{Part11}
Parteli, E., Andrade~Jr, J., and Herrmann, H. (2011).
\newblock Transverse instability of dunes.
\newblock {\em Physical Review E}, 107:188001.

\bibitem[Reffet et~al., 2010]{Reff10}
Reffet, E., Courrech~du Pont, S., Hersen, P., and Douady, S. (2010).
\newblock Formation and stability of transverse and longitudinal sand dunes.
\newblock {\em Geology}, 38:491--494.

\bibitem[Rozier and Narteau, 2014]{Rozi14}
Rozier, O. and Narteau, C. (2014).
\newblock A real-space cellular automaton laboratory.
\newblock {\em Earth Surface Processes and Landforms}, 39:98--109.

\bibitem[Rozier et~al., 2020]{Rozi19}
Rozier, O., Narteau, C., Gadal, C., Claudin, P., and Courrech~du Pont, S.
  (2020).
\newblock Elongation and stability of a linear dune.
\newblock {\em Geophysical Research Letters}, 46(24):14521--14530.

\bibitem[Rubin and Hunter, 1987]{Rubi87}
Rubin, D. and Hunter, R. (1987).
\newblock Bedform alignment in directionally varying flows.
\newblock {\em Science}, 237:276--278.

\bibitem[Telfer et~al., 2018]{Telf18}
Telfer, M., Parteli, E., Radebaugh, J., Beyer, R., Bertrand, T., Forget, F.,
  Nimmo, F., Grundy, W., Moore, J., Stern, S.~A., et~al. (2018).
\newblock Dunes on pluto.
\newblock {\em Science}, 360:992--997.

\bibitem[Valance, 2011]{Vala11}
Valance, A. (2011).
\newblock Nonlinear sand bedform dynamics in a viscous flow.
\newblock {\em Physical Review E}, 83(3):036304.

\bibitem[Ward et~al., 1985]{Ward85}
Ward, A., Doyle, K., Helm, P., Weisman, M., and Witbeck, N. (1985).
\newblock Global map of eolian features on mars.
\newblock {\em Journal of Geophysical Research: Solid Earth}, 90:2038--2056.

\bibitem[Wasson and Hyde, 1983]{Wass83}
Wasson, R. and Hyde, R. (1983).
\newblock Factors determining desert dune type.
\newblock {\em Nature}, 304(5924):337.

\bibitem[Worman et~al., 2013]{Worm13}
Worman, S., Murray, A., Littlewood, R., Andreotti, B., and Claudin, P. (2013).
\newblock Modeling emergent large-scale structures of barchan dune fields.
\newblock {\em Geology}, 41:1059--1062.

\bibitem[Zhang et~al., 2012]{Zhan12}
Zhang, D., Narteau, C., Rozier, O., and {Courrech du Pont}, S. (2012).
\newblock Morphology and dynamics of star dunes from numerical modelling.
\newblock {\em Nature Geoscience}, 5:463--467.

\end{thebibliography}


\newpage

\end{document}